\documentclass[12pt]{iopart}

\usepackage{iopams}  
\usepackage{soul}

\usepackage{bm}
\usepackage{xcolor}
\usepackage{mathrsfs}
\usepackage[utf8]{inputenc}
\usepackage{enumerate}
\usepackage{appendix}
\usepackage{dcolumn}
\usepackage{multirow}
\usepackage{float}

\begin{document}

\title[The type III stress-energy tensor]{The type III stress-energy tensor: Ugly Duckling of the Hawking--Ellis classification}

\author{Prado Mart\'{\i}n--Moruno$^1$ and Matt Visser$^2$}

\address{$^1$Departamento de F\'isica Te\'orica and IPARCOS, \\Universidad Complutense de Madrid, E-28040 Madrid, Spain}
\address{$^2$School of Mathematics and Statistics, Victoria University of Wellington, \\ PO Box 600, Wellington 6140, New Zealand}
\ead{pradomm@ucm.es, matt.visser@sms.vuw.ac.nz} 
\vspace{10pt}
\begin{indented}
\item[]\today
\end{indented}

\begin{abstract}
We present some advances in the understanding of type~III stress-energy tensors as per the Hawking--Ellis classification.
Type~I and type~II naturally appear in classical situations, and can also describe semiclassical effects. Type~IV often shows up in semiclassical gravity. Type~III is much more subtle. 
We focus our attention on type III$_0$ stress-energy tensors, which capture the essence (``essential core'') of type~III. Reflecting on known purely phenomenological examples, (``gyratons''), we are able to generalize the geometry generated by those type~III$_0$ stress-energy tensors.
Moreover, we also succeed in extending work by Griffiths based on massless Weyl spinors by finding a fundamental classical bosonic Lagrangian description of these type III$_0$ stress-energy tensors. To the best of our knowledge this is the first time in the literature that a consistent classical bosonic Lagrangian formulation for type III$_0$ stress-energy has been found.
\end{abstract}

%\pacs{04.20.-q, 04.20.Cv, 04.40.Nr, 04.90.+e, 03.30.+p}

\vspace{2pc}
\noindent{\it Keywords}: type III stress-energy; type III$_0$ stress-energy; 
Hawking--Ellis classification; 
gyratons; Kundt spacetimes. 

\maketitle

\def\tr{{\mathrm{tr}}}
\def\cof{{\mathrm{cof}}}
\def\pdet{{\mathrm{pdet}}}

% For two-column output uncomment the next line and choose [10pt] rather than [12pt] in the \documentclass declaration
%\ioptwocol
%

\section{Introduction}
The understanding of gravity as a property of spacetime shook up the very basis of our understanding of fundamental physics. 
Gravity is related to the spacetime curvature, which dictates the movement of matter and is in turn generated by matter itself. 
In Einstein gravity (general relativity, GR), where no additional degrees of freedom come into play, this relation is established by the proportionality of the Einstein tensor to the stress-energy tensor (SET) associated with the matter content, that is,  the Einstein equations.
Therefore, a detailled understanding the characteristics of SETs is crucial to grasping the properties of physical geometries.

Consider the properties of a symmetric tensor defined in a manifold with Lorentzian signature.
The Hawking--Ellis classification~\cite{H&E,Martin-Moruno:essential} of SETs (also known as the Segr\'e--Pleba\'nski classification)  is based on considering the partial diagonalization of these SETs by local Lorentz transformations. (See also reference~\cite{Martin-Moruno:Rainich} for a different kind of classification in terms of the Rainich conditions.)\footnote{Note that here we are considering a classification of the SETs. We are \emph{not} studying classifications
of the Riemann tensor such as the Petrov classification, which is based on analyzing the algebraic structure of the curvature tensor. 
In particular type~III Petrov has nothing to do with type~III Hawking--Ellis. However, as we will discuss, the essential cores of this Hawking--Ellis classification \cite{Martin-Moruno:essential} do have a direct influence on geometric properties.}
These local Lorentz transformations relate the physics of local observers and, therefore, the partially diagonalized SETs can readily be expressed in an orthonormal basis where we can recognize the energy density, pressures, fluxes, and stresses measured by a comoving observer, connecting with the physics of the problem.
The Hawking--Ellis classification is closely related to the study of the Lorentz invariant eigenvalues and the Lorentz covariant eigenvectors of the mixed tensor $T^a{}_b=T^{ac}g_{cb}$ \cite{Martin-Moruno:LNP}.
The SET of most physical systems we find in nature is of type~I. This class contains $T^a{}_b$'s that have one timelike and 3 spacelike eigenvectors. So, the tensor is completely diagonalizable by means of Lorentz transformations, and the observer comoving with the corresponding orthonormal basis
measures neither fluxes nor stresses.
Type~II SETs have a double null eigenvector. The observer in the orthonormal basis measure a flux in one space direction with the flux given by $f=(\rho+p)/2$, where $\rho$ is the energy density and $p$ the pressure on that direction measured by the observer.
The most relevant physical case described by a type II SET is that corresponding to classical radiation or null dust.
Type~IV SETs have no causal eigenvectors and can be understood as a complex extension of type I \cite{Martin-Moruno:Rainich}.
In this case the orthonormal observer measures a non-zero flux $f$ in one space direction for which $p=\rho$. Known examples of  type~IV are related to semi-classical effects~\cite{Martin-Moruno:LNP,Martin-Moruno:2013sfa,Martin-Moruno:semiclassical,Martin-Moruno:semiclassical-nonlinear}. 

%\clearpage
The only remaining type in the Hawking--Ellis classification is type~III. Type~III SETs have a triple null eigenvector and the orthonormal observers not only measure fluxes but also stresses.
In particular, in one of the 2 spatial directions in which the observer necessarily measures $p=-\rho$ she sees a flux $f$, and in the other she measures a stress with respect to the former direction also with value $f$.
From a physical point of view, this appears to be the most peculiar of the SET types, as examples in (classical or semi-classical) situations are very difficult to find.
In view of this it seems reasonable to refer to type III SETs as the ``ugly ducking'' of the Hawking--Ellis classification.

An extension of the Hawking--Ellis classification to arbitrary dimensions can be found in reference~\cite{Maeda:2018}.
However, note that the formulation of the type III SET that we are using, based on the analysis in reference~\cite{Martin-Moruno:LNP}, 
is not unnecessarily restrictive as is claimed in reference~\cite{Maeda:2018}.
The SET considered by those authors was 
\begin{equation}\label{SETMaeda}
T_{ab}=-\rho g_{ab}+f_1k_ak_b+f_2(k_as_b+s_ak_b)+p_3n_an_b;
\end{equation}
 with $k^2=0$, $s^2=1$ and $k\cdot s=0$. But this can also be expressed as 
\begin{equation}\label{typeIIISET}
T_{ab}=-\rho g_{ab}+f_2(k_a\tilde s_b+\tilde s_ak_b)+p_3n_an_b,
\end{equation}
where we now define $\tilde s_a=\left[s_a+f_1/(2f_2)\,k_a\right]$, which corresponds\,\footnote{One can explicitly show that (\ref{SETMaeda}) and (\ref{typeIIISET}) are related by a Lorentz transformation by first performing a rotation in the $x-y$ plane, with (real) angle $\theta$ fixed by $\cos\theta=\zeta/\sqrt{\zeta^2+(\nu/2)^2}$, and then a (subluminal) boost in the $y$ direction, with $\beta=(\nu/2)/\sqrt{\zeta^2+(\nu/2)^2} <1$.}  to the form presented in reference~\cite{Martin-Moruno:LNP}.

%\enlargethispage{45pt}
In order to advance in the understanding of type III SETs, we have first defined the essential cores of the type II--IV tensors \cite{Martin-Moruno:essential}, those are types II$_0$--IV$_0$. 
Those essential cores are intended to capture the essence of the corresponding SET types.
They are defined by subtracting out from the SET as much as possible of type~I, which simplifies the eigenvalue structure, while preserving the eigenvector structure. 
In addition they have the interesting property that, in standard general relativity, they generate geometries in which the Ricci tensor is of the same type as  the SET.
(Since $T\equiv T^a{}_a=0$, then $R=0$ and $R^a{}_b\propto T^a{}_b$.)
We have reported preliminary advances in understanding the physics of type~III SETs in reference \cite{Martin-Moruno:III_0}.
Although we found some specific examples type III$_0$ geometries, and explored the Lagrangian formulation of the corresponding matter content, we did not find as clear a physical intuition as we may have wished for those geometries,
and failed to extend our Lagrangian analysis beyond the flat spacetime. In the present work, we go beyond our previous study,  taking care of those limitations.

This paper can be outlined as follows: In section (\ref{sec:gyraton}) we review part of the literature, pointing out a phenomenological example of type III$_0$ SET that will be useful for our study.
In section (\ref{sec:geometry}) we generalize the metric of this phenomenological example as much as we can, to find a quite general type III$_0$ geometry. 
In section (\ref{sec:Lagrangian}) we comment on a known example in the literature of a fundamental description of Bonnor's gyraton, showing what constructions could work when looking for a type III$_0$ Lagrangian, and those that could not, pointing out two particular examples.
Finally, in section (\ref{sec:conclusions}) we summarize our conclusions.

%========================================================
\section{Phenomenological model: The gyraton}\label{sec:gyraton}
%========================================================

As all essential core types, II$_0$, III$_0$ and IV$_0$ are traceless~\cite{Martin-Moruno:essential}, and as in this work we assume the validity of standard general relativity, 
we shall temporarily focus our attention on vanishing scalar invariant spacetimes.
(However, note that this condition is more restrictive than just requiring $R=0$.) 
These vanishing scalar invariant spacetimes are a subclass of Kundt's geometries \cite{Pravda:2002us}.
Per definition a Kundt spacetime is a geometry having a null vector that is geodesic, expansion-free, shear-free and twist-free. 
Taking $\ell=\partial_r$, and ordering our coordinates as $x^a=(u,r,x^i)$, the metric of Kundt spacetimes can be written in the reasonably standard form \cite{Stephani,Griffiths,Podolsky:2008,Coley:2009}
\begin{equation}\label{Kundt}
\fl\qquad\qquad
 ds^2= -2 du \left[dr-H(u,\,r,\,x^k)\,du-W_i(u,\, r,\,x^k)\,dx^i\right]+g_{ij}(u\,,x^k)\,dx^idx^j.
\end{equation}
All the spacetimes directly studied in this work are four-dimensional. (We do sometimes refer back to some (2+1)-dimensional examples previously discussed in the literature.)
This spacetime depends on 6 different arbitrary functions of $u$, $r$, and $x^i=(x,y)$.  Note specifically that there is no $r$ dependence in the $g_{ij}(u\,,x^k)$. 
It is also useful to have at hand the explicit form of the matrix of metric and inverse-metric coefficients:
\begin{equation}
\fl\qquad
g_{ab} = \left[ \begin{array}{c|c|c} 2 H & -1 &  W_j \\ \hline -1 & 0 & 0\\ \hline  W_i & 0 & g_{ij}\end{array} \right];
\qquad
g^{ab} = \left[ \begin{array}{c|c|c}  0& -1 &  0  \\ \hline -1 & -2H+g^{kl}W_kW_l &g^{jk} W_k\\ \hline  0 & g^{ik} W_k & g^{ij}\end{array} \right].
\end{equation}
Here $g^{ij}$ is the matrix inverse of the nonsingular $2\times2$ matrix $g_{ij}$. 
It is worth noting that $r$ is typically a non-null coordinate (so that $(dr)_a$ is a non-null covariant vector with $g^{rr}\neq0$), but $r$  is a null parameter (in the sense that $(\partial_r)^a$ is a null contravariant vector with $g_{rr}=0$). 
In contrast $u$ is a null  coordinate (so that $(du)_a$ is a null covariant vector with $g^{uu}=0$), but $u$  is typically a non-null parameter (in the sense that $(\partial_u)^a$ is a non-null contravariant vector with $g_{uu}\neq0$).
See reference~\cite{Coll:2009zza} for some comments about the subtlety of associating unambiguous causal character to  coordinates in general situations.

Geometries generated by a type III$_0$ SET have $R_{ab}\neq0$, $(R^2)_{ab}\neq0$ and $(R^3)_{ab}=0$.
Note that we already know that they cannot be spherically symmetric nor even planar symmetric~\cite{Martin-Moruno:essential}.
In our literature search for examples, apart from our own recent examples in reference~\cite{Martin-Moruno:III_0}, we only found the \emph{gyraton}.
These gyratons are purely phenomenological models intended to describe the gravitational field of a localized ultrarelativistic source with an intrinsic rotation moving at the speed of light.
In 1970 Bonnor already showed  \cite{Bonnor} that the SET of this phenomenological spinning null fluid satisfies $(T^3)_{ab}=0$, with $(T^2)_{ab}\neq0$ 
if the angular momentum is non-vanishing. Therefore, it is of type III$_0$. However, Bonnor did not provide any fundamental Lagrangian description to derive such a SET.
The characteristics of these gyratons on various (3+1)-dimensional background spacetimes are summarized in the relatively recent reference~\cite{Kadlecova:2013uta,Kadlecova:2009qu}. 
(See also the more recent~\cite{Podolsky:2018zha} for (2+1)-dimensional examples,
\cite{Podolsky:2014lpa} for a review and physical analysis of $pp$-wave gyratons in $4$ dimensions, and
\cite{Podolsky:2018oov} for a very recent work in which the full class of gyratons in the Robinson-Trautman and Kundt classes were derived.)
Note that the term ``gyraton'' was not introduced by Bonnor himself, but it seems to have its origin in references \cite{Frolov:2005in,Frolov:2005zq,Frolov:2005ww}. 

%\clearpage
%\enlargethispage{35pt}
\paragraph{\bf Example 1:}
%{\bf Example 1:}  
The simplest form of the gyraton in a 4-dimensional spacetime is then that of the $pp$-wave Brinkmann metric which is most commonly written in the form \cite{Griffiths,Kadlecova:2013uta}
\begin{equation}\label{Brinkmann}
 ds^2= -2 du \,dr+\Phi\left(u,\,x^i\right)du^2 +(dx^i)^2+2a_i\left(u,\,x^i\right)\,dx^i\,du.
\end{equation}
Here $\Phi$ and $a_i$ are known as the gravitoelectric and gravitomagnetic potential, respectively.
For this form of the $pp$-wave Brinkmann metric one has
\begin{equation}
g_{ab} = \left[ \begin{array}{c|c|c} 
\Phi & -1 &  a_j \\ \hline 
-1 & 0 & 0\\ \hline  
a_i & 0 & \delta_{ij}
\end{array} \right];
\qquad
g^{ab} = \left[ \begin{array}{c|c|c}
  0& -1 &  0  \\ \hline 
  -1 & -\Phi+\delta^{kl}a_ka_l & \;\;\delta^{jk} a_k\; \; \\ \hline  
  0 & \; \delta^{ik} a_k \; & \delta^{ij}
  \end{array} \right].
\end{equation}
The use of multiple slightly different notations and sign conventions is unfortunately quite common.
This spacetime has  a null vector field $\ell = \partial_r$ for which
\begin{equation}
\ell^a=(0,1,0,0)\qquad\hbox{so} \qquad \ell_a=(-1,0,0,0).
\end{equation}
Here we can write the covector $\ell_a$ as $\ell_a = - \partial_a u$. 
This geometry has, therefore, a covariantly constant null vector, that is  $\nabla_a\ell_b=0$, which implies that the rays of the non-expanding waves will be parallel and the wave fronts planar. 
(As discussed in the appendix, the condition of being shear-free and non-expanding already implies $\nabla_i\ell_j=0$.)\\
The only nonzero components of the Ricci tensor are
\begin{equation}
\fl
\qquad\qquad
R_{uu} = -{\partial_x^2\Phi+\partial_y^2\Phi\over2} +\partial_u(\partial_x a_x + \partial_y a_y) + {f^2\over2}; \qquad 
R_{ui} = \frac{1}{2}\,( \partial_y f, -\partial_x f);
\end{equation}
where $f= \partial_x a_y - \partial_y a_x$. Consequently $R=0$, while
\begin{equation}
(R^2)_{ab} =\frac{1}{4}\, \{(\partial_x f)^2 + (\partial_y f)^2\} \, \ell_a \ell_b,
\end{equation}
and $(R^3)_{ab}=0$. 

In what sense is this geometry ``rotating''? 
Consider the $(x,y)$ plane, transverse to the $(u,r)$ plane. In the $(x,y)$ plane the metric is particularly simple: $g_{ij} = \delta_{ij}$. 
Then the angular momentum per unit length around the point $(x,y)=(0,0)$ can be defined as
\begin{equation}
  J(u)  = \int  \left[ (x,y) \times (T_{ux}, T_{uy}) \right] dx dy,
\end{equation}
  that is
\begin{equation}
   J(u) = \int  \left[ x T_{uy} -  y T_{ux} \right] dx dy. 
\end{equation}
For our purposes, using the Einstein equations, we see
\begin{equation}
   8 \pi G_N J(u) = \int  [ x R_{uy} -  y R_{ux} ] dx dy.
\end{equation}
Now for our example 1 above we see
\begin{eqnarray}
\int  \left[ x R_{uy} -  y R_{ux} \right] dx dy  &=& -{1\over2} \int  \left[ x \partial_x f +  y \partial_y f  \right] dx dy 
\nonumber\\
   &=& -{1\over2}  \int  \left[ \partial_x  (xf) +  \partial_y (yf)  \right] dx dy  + \int  f  dx dy.
\end{eqnarray}
With suitable falloff conditions at large $|x|$ and $|y|$, the Gauss theorem lets us discard the first (pure divergence) term so
\begin{equation}
8\pi G_N  J(u) =   \int  f \; dx dy = \int \left(\partial_x a_y - \partial_y a_x\right) dx dy.
\end{equation}
Note that final result is independent of where you put the origin $(0,0)$ of the $(x,y)$ plane.
Ultimately, this non-zero angular momentum is what justifies calling the ``gyraton'' an example of a ``spinning'' null fluid.
We shall now seek to both generalize and specialize this example in several ways.

For instance, for the case of gyraton propagation in a direct-product spacetime filled with an electromagnetic field, the metric takes the slightly more general form \cite{Kadlecova:2013uta}
\begin{equation}\label{Kadlecova}
\hspace{-1.5cm} ds^2= -2\, du \,dr+\Phi\left(u,\,r,\,x^i\right)du^2 +\frac{1}{P(x^i)^2}\left[\delta_{ij} \,dx^i \,dx^j\right]+2a_i\left(u,\,x^i\right)\,dx^idu,
\end{equation}
where in reference \cite{Kadlecova:2013uta} the author focuses on the case
\begin{equation}\label{caseK}
P(\partial_x^2P+\partial_y^2P)-(\partial_xP)^2-(\partial_yP)^2=\Lambda_+, \quad {\rm and}\quad \partial_r^2\Phi=\Lambda_-,
\end{equation}
with $\Lambda_+$ and $\Lambda_-$ being constants.
%\enlargethispage{20pt}
It can easily be verified that this geometry generically does not have a type III$_0$ Ricci tensor, since $R=2\Lambda_+ + \Lambda_-$ which is nonzero in general (see next section).
Nor does this geometry even have a general type~III Ricci tensor. (Generically there are 2 different eigenvalues of $R^a{}_b$, namely $\Lambda_+$ and $\Lambda_-$, and both have multiplicity 2.)
Moreover, this spacetime does not possess a covariantly constant null vector, since one now has $\nabla_a \ell_b = (\partial_r \Phi) \, \ell_a\ell_b$ which is non-zero (see appendix).\\
The gyraton source is again described only at a purely phenomenological level. (Some authors have suggested that the gyraton SET  could be obtained from an approximation to an electromagnetic Lagrangian \cite{Frolov:2005in}.
We will discuss that issue in section \ref{sec:Lagrangian} in more detail.)

\paragraph{\bf Example 2:}
%{\bf Example 2:} 
Observe that a mildly interesting \emph{specialization} of the spacetime presented in equation (\ref{Brinkmann}) is to take the particularly simple case
\begin{equation}\label{simple}
 ds^2= -2 du \,dr +(dx^i)^2+2a_i(x^i)\,dx^i\,du.
\end{equation}
The only nonzero components of the Ricci tensor are now
\begin{equation}
R_{uu} =  {f^2\over2}; \qquad R_{ui} = \frac{1}{2} ( \partial_y f; -\partial_x f);
\end{equation}
where again $f= \partial_x a_y - \partial_y a_x$. Consequently $R=0$, while
\begin{equation}
(R^2)_{ab} =\frac{1}{4}\,\{(\partial_x f)^2 + (\partial_y f)^2\}\, \ell_a \ell_b,
\end{equation}
and $(R^3)_{ab}=0$. 
As in example 1, for the angular momentum per unit length we still have the simple formula
\begin{equation}
8\pi G_N  J(u) =   \int  f \; dx dy = \int \left(\partial_x a_y - \partial_y a_x\right) dx dy.
\end{equation}

%====================================================
\section{General type III$_0$ geometry}\label{sec:geometry}
%====================================================

Now consider the following increasingly general examples of type III$_0$ geometry:

\paragraph{\bf Example 3:}
%{\bf Example 3:}  
%Let now us 
We now study geometry (\ref{Kadlecova}) in more detail to try to modify it to find a general type III$_0$ spacetime.
As already noted, this metric does not generically have vanishing scalar invariants. 
However, explicitly calculating the scalar curvature we obtain
\begin{equation}\label{R}
R=2P\left[\partial_x^2P+\partial_y^2P\right]-2\left[\left(\partial_xP\right)^2+\left(\partial_yP\right)^2\right]+\partial_r^2\Phi,
\end{equation}
which, taking into account conditions (\ref{caseK}), reduces to $R=2\Lambda_++\Lambda_-$.
So, the Ricci scalar vanishes if we take $2\Lambda_+=-\Lambda_-$. A particularly simple case satisfying this $R=0$ condition is
\begin{equation}\label{c}
P={\rm constant}\quad{\rm and}\quad \Phi\left(r,\,u,\,x^i\right)\quad {\rm with}\quad \partial_r^2\Phi=0.
\end{equation}
We have explicitly checked that under these conditions $(R^3)_{ab}=0$, and that the only non-vanishing component of $(R^2)_{ab}$ is 
\begin{equation}\label{Ruu}
(R^2)_{uu} =\frac{1}{4}P^2\left[( \partial_r\partial_y\Phi+P^2\partial_xf)^2+(\partial_r\partial_x\Phi-P^2\partial_yf)^2\right].
\end{equation}
Here we have defined $f(u,x,y) = \partial_x a_y(u,x,y) - \partial_y a_x(u,x,y)$.

Now, after restricting to the $R=0$ case, we can make our first generalization beyond (\ref{Kadlecova}) and consider a non-constant $P=P(u)$. 
(Note that we are not particularly interested in having all spacelike 2-surfaces with the same value of curvature. However, that curvature should be independent of $r$ to have a non-expanding congruence of null geodesics, see the appendix.) That is, we now set
\begin{equation}\label{Kadlecova_u}
\hspace{-1cm} ds^2= -2\, du \,dr+\Phi\left(u,\,r,\,x^i\right)du^2 +\frac{1}{P(u)^2}\left[\delta_{ij} \,dx^i \,dx^j\right]+2a_i\left(u,\,x^i\right)\,dx^idu.
\end{equation}
If we keep $\partial_r^2\Phi=0$, we will still have $\nabla_a \ell_b = (\partial_r \Phi) \, \ell_a\ell_b$, even if it is not necessarily covariantly constant, while both $R=0$ and 
$(R^3)_{ab}=0$. The only non-vanishing component of $(R^2)_{ab}$ given by equation (\ref{Ruu}), but with $P=P(u)$. 
For completeness we note that the only non-zero components of the Ricci tensor are now
%===========
\begin{eqnarray}
R_{uu} &=&-  {1\over2} P^2\delta^{ij} \,\partial_i \partial_j \Phi + 2{\partial_u^2 P\over P} -4  {(\partial_u P)^2\over P^2} - P^2 \delta^{ij} a_i \partial_j \partial_r\Phi
-{\partial_r \Phi \, \partial_u P\over P}
\nonumber\\
&& 
+(P^2\delta^{ij} \, \partial_u a_{i,j}) - {1\over2} \partial_r\Phi (P^2\delta^{ij}  a_{i,j})+ {1\over2} {P^4f^2},
\end{eqnarray}
%===========
and
\begin{equation}
R_{ui} =  {1\over2}\left(-\partial_x \partial_r \Phi + P^2 \partial_y f; \; -\partial_y \partial_r \Phi - P^2 \partial_x f  \right)_i.
\end{equation}
That is, temporarily defining $s_a = (0,0; R_{ui})$ we have $\ell^a s_a=0$ and the Ricci tensor takes the form
\begin{equation}
R_{ab} = (R_{uu}) \ell_a \ell_b + \ell_a s_b + s_a \ell_b.
\end{equation}
This is enough to guarantee 
that one has 
\begin{equation}
(R^2)_{ab} = (g^{cd} s_c s_d)\, \ell_a \ell_b = (P^2 \delta^{ij} R_{ui} R_{uj} )\, \ell_a \ell_b,
\end{equation}
thereby implying $(R^3)_{ab}=0$.
Thence, in order to have a type III$_0$ Ricci tensor, we need either 
\begin{equation}\label{conditions}
 \partial_r\partial_ y\Phi + P^2 \partial_xf \neq0  \qquad{\rm or} \qquad \partial_r\partial_x\Phi - P^2 \partial_yf\neq0.
\end{equation}
Note that for $\partial_xf=\partial_yf=0$ and $\partial_r\partial_x\Phi=0=\partial_r\partial_y\Phi$ the Ricci tensor of this spacetime reduces to type II$_0$.

The calculation of the angular momentum per unit length is now slightly modified, though the ultimate result remains the same. 
In the $(x,y)$ plane we first note $g_{ij}= P^{-2} \delta_{ij}$ so for the area element $dx dy \longrightarrow P^{-2} dx dy$. 
However when calculating the angular momentum per unit length we must also consider what happens to the integrand $\left[ (x,y) \times (T_{ux}, T_{uy}) \right]$.
To see what happens it is best to write the original version of the $J$-integrand as
\begin{equation}
\hspace{-1cm}(\hbox{$J$-integrand}) = \left[ (x,y) \times (T_{ux}, T_{uy}) \right] = (\epsilon_0)^{ij}  (\delta_{ik} x^k) T_{uj} = (\epsilon_0)_{ij} \;x^i \;(\delta^{jk} T_{uk}).
\end{equation}
Here $\epsilon_0$ is the Cartesian Levi-Civita 2-tensor, normalized to $(\epsilon_0)_{xy}=+1$. \\
The generalization to $g_{ij}\neq\delta_{ij}$ is now obvious. Take the integrand to be
\begin{equation}
(\hbox{$J$-integrand}) = \epsilon^{ij}  (g_{ik} x^k) T_{uj} = \epsilon_{ij} \;x^i \;(g^{jk} T_{uk}).
\end{equation}
Then we see that, as we switch on $P(u)$, we have $\epsilon^{ij} \to P(u)^{2} (\epsilon_0)^{ij}$ and $g_{ik} \to P(u)^{-2} \delta_{ij}$, so that the $J$-integrand is left  undisturbed.
So we now have
\begin{equation}
   8 \pi G_N J(u) = P(u)^{-2} \int \left[ x R_{uy} -  y R_{ux} \right] dx dy .
\end{equation}
But now there is an extra contribution to the integral. We have
\begin{eqnarray}
\fl
P(u)^{-2} \int  \left[ x R_{uy} -  y R_{ux} \right] dx dy  &=& -{1\over2} \int  \left[ x \partial_x f +  y \partial_y f  \right] dx dy 
\nonumber\\
\fl
&& -{P(u)^{-2}\over2} \int  \left[ x \partial_y \partial_r\Phi +  y \partial_x \partial_r \Phi  \right] dx dy.
\end{eqnarray}
The new contribution to the integral is 
\begin{equation}
 \int  \left[ x \partial_y \partial_r\Phi +  y \partial_x \partial_r \Phi  \right] dx dy  
= \int  \left[ \partial_x (y\partial_r \Phi)+ \partial_y (x \partial_r\Phi)   \right] dx dy .
\end{equation}
With suitable falloff conditions at large $|x|$ and $|y|$, the Gauss theorem lets us discard this term. Note that this term is automatically absent in any metric with $\partial_r\Phi=0$ 
(as, for example in \cite{Frolov:2005in}) and, therefore, the falloff condition will be satisfied.
So we still have
\begin{equation}
8\pi G_N  J(u) =   \int  f \; dx dy = \int \left(\partial_x a_y - \partial_y a_x\right) dx dy.
\end{equation}
Note that final result is again independent of where you put the origin $(0,0)$ of the $(x,y)$ plane.

\paragraph{\bf Example 4:}
%{\bf Example 4:}
Metric (\ref{Kadlecova_u}) can also be \emph{specialized} in an interesting manner. Consider
\begin{equation}\label{special2}
 ds^2= -2\, du \,dr+\Phi\left(u,\,r,\,x^i\right)du^2 + \delta_{ij}\, dx^i dx^j,
\end{equation}
where $\partial_r^2\Phi=0$. 
Then we have the very simple results that 
\begin{equation}
R_{uu} =  - {1\over2} \delta^{ij} \,\partial_i \partial_j \Phi;    \qquad R_{ui} = - {\partial_i \partial_r \Phi\over2}.
\end{equation}
Then $R=0$, $(R^3)_{ab}=0$ and
\begin{equation}
(R^2)_{ab} = {1\over4}\left\{ \delta^{ij} (\partial_i \partial_r \Phi)(\partial_j \partial_r\Phi) \right\} \ell_a \ell_b.
\end{equation}
This again is a simple form of type III$_0$, as long as one of the $\partial_i \partial_r \Phi$ is non-zero.

Note that for this particular geometry the (integrated) angular momentum is zero: $J(u)\to 0$; so it is not ``spinning'' overall but is still certainly type III$_0$.
For this specific example the $J$-integrand need not be zero, except in cases of cylindrical symmetry. In the absence of cylindrical symmetry the $J$-integrand $\neq 0$, but the integrated value of $J=0$. 
Thus this example proves that $J\neq 0$ is not necessary for a type III$_0$ SET. 
Certainly $J\neq0$ implies $T_{ui}\neq0$ which is related to type III$_0$ behaviour. But $J=0$ still permits $T_{ui}\neq0$, and type III$_0$ behaviour, albeit in a restricted manner.

It is easy to compare this specialized metric (\ref{special2}) with the (2+1)-dimensional example of reference~\cite{Martin-Moruno:III_0}. One need only reduce by one spatial dimension, then choose $\Phi=2\sqrt{2}\,xrf$, and set $\delta_{ij}\to 1$.

\paragraph{\bf Example 5:}
%{\bf Example 5:} 
Let us now further generalize metric (\ref{Kadlecova_u}), while still restricting ourselves to the $R=0$ case, by looking for spacetimes that still have a type III$_0$ Ricci tensor but that are not necessarily product spacetimes. A more general metric, still of the Kundt class, that we have found is this:
\begin{equation}\label{Km}
 ds^2= -2\, du \,dr+\Phi\left(u,\,r,\,x^i\right)du^2 +\frac{1}{P_i^2\left(u\right)}(dx^i)^2+2a_i\left(u,\,x^i\right)\,dx^idu.
\end{equation}
For $\partial_r^2\Phi=0$ this geometry has $R=0$ and $(R^3)_{ab}=0$, while 
\begin{equation}
R_{ui} =- {1\over2} (\partial_r\partial_x\Phi , \partial_r\partial_y\Phi )+  {1\over2} (P_y^2\,\partial_y f, - P_x^2 \partial_x f).
\end{equation}
The only non-vanishing component of $(R^2)_{ab}$ is
\begin{equation}
\hspace{-1cm}(R^2)_{uu} = g^{ij} \,R_{ui} \,R_{uj} =\frac{1}{4}\left\{P_y^2(P_x^2\partial_xf+\partial_r\partial_y\Phi)^2+P_x^2(-P_y^2\partial_yf+\partial_r\partial_x\Phi)^2\right\}.
\end{equation}
So, since $(R^2)_{uu}$ is a sum of squares,  at least one of conditions
\begin{equation}\label{conditions2}
 \partial_r\partial_y\Phi + P_x^2 \partial_xf \neq0  
\qquad{\rm or} \qquad 
\partial_r\partial_x\Phi - P_y^2 \partial_yf\neq0.
\end{equation}
has to hold to avoid a type II$_0$ Ricci tensor.
Note that for  $\partial_r^2\Phi\neq0$ the Ricci tensor is also not a type III tensor, as $R^a{}_b$ has 2 eigenvalues with multiplicity 2, namely 0 and  $\partial_r^2\Phi$.

\def\diag{{\mathrm{diag}}}

The calculation of the angular momentum per unit length is again slightly modified, though the ultimate result remains the same. 
In the $(x,y)$ plane we first note $g_{ij}= \diag\{P_x^{-2},P_y^{-2}\}_{ij}$ so for the area element $dx dy \longrightarrow (P_x P_y)^{-1} dx dy$. 
However when calculating the angular momentum per unit length we must also note
\begin{equation}
(\hbox{$J$-integrand}) = \epsilon^{ij}  (g_{ik} x^k) T_{uj} = \epsilon_{ij} \;x^i \;(g^{jk} T_{uk}).
\end{equation}
Then we see that, as we switch on the $P_i(u)$, we have $\epsilon^{ij} \to (P_x(u)P_y(u))\;(\epsilon_0)^{ij}$ and since $g_{ij}\to \diag\{P_x^{-2},P_y^{-2}\}_{ij}$
the $J$-integrand transforms as
\begin{equation}
(\hbox{$J$-integrand}) \to {P_y\over P_x} \; x T_{uy} - {P_x\over P_y} \; y T_{ux}. 
\end{equation}
So now we have
\begin{equation}
   8 \pi G_N J(u) = \int  \left[ {x R_{uy}\over P_x^2} -  {y R_{ux}\over P_y^2} \right] dx dy .
\end{equation}
But now there is an extra contribution to the integral
\begin{eqnarray}
\int  \left[ {x R_{uy}\over P_x^2} -  {y R_{ux}\over P_y^2} \right] dx dy  &=& -{1\over2} \int  \left[ x \partial_x f +  y \partial_y f  \right] dx dy 
\nonumber\\&&
-{1\over2} \int  \left[ {x \partial_y \partial_r\Phi\over P_x^2} +  {y \partial_x \partial_r \Phi\over P_y^2}  \right] dx dy .
\end{eqnarray}
The new contribution to the integral is 
\begin{equation}
\int  \left[ {x \partial_y \partial_r\Phi\over P_x^2} +  {y \partial_x \partial_r \Phi\over P_y^2}  \right] dx dy  
= \int  \left[ \partial_y \left(x \partial_r\Phi\over P_x^2\right) +  \partial_x \left(y\partial_r \Phi\over P_y^2\right)  \right] dx dy .
\end{equation}
With suitable falloff conditions at large $|x|$ and $|y|$, the Gauss theorem lets us discard this term so we still have
\begin{equation}
8\pi G_N  J(u) =   \int  f \; dx dy = \int \left(\partial_x a_y - \partial_y a_x\right) dx dy.
\end{equation}
Note that final result is again independent of where you put the origin $(0,0)$ of the $(x,y)$ plane.

\paragraph{\bf Example 6:}
%{\bf Example 6:}
Metric (\ref{Km}) can be generalized even further. The most general metric we have found with a type~III$_0$ Ricci tensor is
\begin{equation}\label{general}
 ds^2= -2\, du \,dr+\Phi\left(u,\,r,\,x^i\right)du^2 +2a_i\left(u,\,x^i\right)dx^i\,du+ g_{ij}(u) dx^i dx^j,
\end{equation}
where $\partial_r^2\Phi=0$ and $g_{ij}$ is a nonsingular $2\times2$ matrix of Euclidean signature.
It is easy to check that for this metric $R=0$ and $(R^3)_{ab}=0$. The only nonzero components of the Ricci tensor are
a little trickier to evaluate. $R_{uu}$ is non-zero, quite messy, and not particularly interesting.
In contrast a relatively clean result is
\begin{equation}
R_{ui} =- {1\over2} \{\partial_r\partial_i\Phi - (\epsilon_0)_{ik} \,g^{kl} \,\partial_l f\}.
\end{equation}
Here $(\epsilon_0)_{ij}$ is again the 2-dimensional Levi--Civita symbol normalized by  $(\epsilon_0)_{xy}=1$. 
We could also write this as
\begin{equation}
R_{ui} =- {1\over2} \left(\partial_r\partial_i\Phi - {\epsilon_{ik}\over\sqrt{\det(g_{ij})}} \,g^{kl} \,\partial_l f\right).
\end{equation}
For $(R^2)_{ab}$ there is only one nonzero component, namely
\begin{equation}
\hspace{-2cm}(R^2)_{ uu} = g^{ij} R_{ui} R_{uj} = \frac{1}{4}\left\{
g^{ij}(\partial_r\partial_i\Phi - (\epsilon_0)_{ik} \,g^{kl} \,\partial_l f)(\partial_r\partial_j\Phi - (\epsilon_0)_{jm} \,g^{{mn}} \, \partial_{n} f)
\right\}.
\end{equation}
In addition, to avoid $(R^2)_{ab}=0$ one needs at least one of the components of the 2-vector
\begin{equation}
\partial_r\partial_i\Phi - (\epsilon_0)_{ik} \,g^{kl} \,\partial_l f
\end{equation}
to be non-zero.
Moreover, we could also reparameterize the $u$ coordinate and include a function $G(u)$ in the term $du\,dv$. 
Note that the null vector $\ell$ still satisfies the condition $\nabla_a\ell_b = (\partial_r\Phi)\, \ell_a \ell_b$.

To calculate the angular momentum per unit length note that the area element is $dx dy \to \sqrt{\det(g_{ij})} \; dx dy$ and that
the relevant $J$-integrand is 
\begin{equation}
(\hbox{$J$-integrand}) = \epsilon^{ij}  (g_{ik} x^k) R_{uj} = -{1\over2}\epsilon^{ij}  (g_{ik} x^k) (\partial_r\partial_j\Phi - (\epsilon_0)_{jm} \,g^{ml} \,\partial_l f).
\end{equation}
Thence 
\begin{equation}
(\hbox{$J$-integrand}) = -{1\over2} {\left(  x^j \partial_j f \right)\over\sqrt{\det(g_{ij})}} - {1\over2} \partial_j (\partial_r\Phi) \epsilon^{ij} (g_{ik} x^k).
\end{equation}
That is
\begin{equation}
(\hbox{$J$-integrand}) = -{1\over2} {\partial_j \left(  x^j  f \right)\over\sqrt{\det(g_{ij})}} + {f\over\sqrt{\det(g_{ij})}} - {1\over2} \partial_j (\partial_r\Phi \; \epsilon^{ij} (g_{ik} x^k)).
\end{equation}
The divergence terms again drop out due to the Gauss theorem and suitable falloff conditions,  so again we have
\begin{equation}
8\pi G_N  J(u) =   \int  f \; dx dy = \int \left(\partial_x a_y - \partial_y a_x\right) dx dy.
\end{equation}
The persistence of this formula for the angular momentum is due to the fact that ultimately one is integrating a 2-form over a 2-plane, 
and that given the form of the underlying spacetime metric this the only plausible term that could arise.

The comparison of metric (\ref{general})  with the (3+1)-dimensional example of reference \cite{Martin-Moruno:III_0} is not straightforward,  as in reference \cite{Martin-Moruno:III_0} the geometry is expressed in Rosen-inspired form, and not in the (now we know) more natural Brinkmann-inspired form. However, one can note that the conditions 
that led to type III$_0$ geometries in those situations also implied the existence of a null Killing vector that is covariantly constant. So, those geometries can (after suitable coordinate transformations) be understood to be a particular subclass of the spacetimes considered in this paper. (See the discussion in the appendix.)
On the other hand, one can compare metric (\ref{general}) with the general (2+1)-Kundt spacetime presented in reference \cite{Podolsky:2018zha}. For this purpose, we first reduce by one spatial dimension $dx^i\rightarrow dx$.
Then, one can easily see that our case corresponds to $\partial_r^2g_{uu}=0$, $\partial_rg_{ux}=0$, and $\partial_ug_{xx}=0$.

%\enlargethispage{30pt}
%========================================================
\section{Fundamental Lagrangian description}\label{sec:Lagrangian}
%========================================================
In the previous section we have discussed the existence of a geometry that can be generated by a type III$_0$ SET (for preliminary ideas along these lines, see reference \cite{Martin-Moruno:III_0}).
As we have briefly discussed, this metric is related with the gyraton.
Nevertheless, in order to go into more depth in understanding this kind of SET, in our opinion,
we should find some fundamental Lagrangian description of the matter source and its state. That is, we want to find a Lagrangian leading to a type III$_0$ SET, in which the matter is not treated merely as a phenomenological source term whose internal structure is unknown.

%========================================================
\subsection{What does not work}
%========================================================
As the gyraton is a particular kind of null fluid, one could be tempted to use the electromagnetic Lagrangian to look for particular solutions with the desired properties. (Such a line of thinking has been followed, for example, in reference \cite{Frolov:2005in}.)
However, as is well-known, the electromagnetic SET satisfies the Rainich condition. That is, (see for example reference~\cite{Martin-Moruno:Rainich}),
\begin{equation}\label{R1}
 (T^2)_{ab} = \frac{1}{4} \, \tr(T^2) \, g_{ab} = \frac{1}{4}\, (T^{cd} \,T_{cd})  \, g_{ab}, \qquad \hbox{with} \qquad \tr(T) =0.
\end{equation}
This condition implies that:
\begin{itemize}
 \item If $\tr(T^2) = T^{cd} T_{cd} = 0$, then  $(T^2)_{ab} =0$ and the EM SET is type II$_0$.
 
 \item If $\tr(T^2)=T^{cd} T_{cd} \neq 0$, then $(T^3)_{ab} \propto T_{ab}\neq 0$, and hence the EM SET is certainly not type III$_0$.
 
\item Moreover, if $\tr(T^2)=T^{cd} T_{cd} \neq 0$, then equation (\ref{R1}) implies that the degree of the minimal polynomial of $T^a{}_b$ is 2. 
(That is the polynomial with the lowest degree such that $m(T^a{}_b)=0$, see reference~\cite{Martin-Moruno:Rainich} for more details.)
This is of the form $m(\lambda)=(\lambda-\lambda_1)(\lambda-\lambda_2)=\lambda^2+\lambda_1\lambda_2$, with $\lambda_1$ and $\lambda_2$ being the eigenvalues, and therefore $\lambda_1=-\lambda_2\neq0$.
As the exponent of both factors in $m(\lambda)$ is 1, then that is the dimension of the largest Jordan blocks; so,  $T^a{}_b$ is diagonalizable and, therefore, it is type I.
 \end{itemize}
That is, at least in 4 dimensions, the electromagnetic stress energy is \emph{never} type III.

In view of this exact result any argument suggesting that a gyraton may be described by an approximate electromagnetic Lagrangian in 4 dimensions has to be misleading. 
Note that only types~I and IV SETs are stable under generic perturbations \cite{Martin-Moruno:essential}.
Indeed, arbitrarily small perturbations will change the very fragile eigenvalue and eigenvector structure required for the existence of types~II and III. 
So, any approximation scheme leading to a type~III SET cannot be trusted. It should be emphasized that we are not suggesting that the study of a class of spacetimes is irrelevant because the algebraic structure of a geometric object is unstable under small perturbations; merely that it cannot be concluded that a SET is type~III by using an approximate procedure. However, it is worthy to note that the nature of the SET is not just an arbitrary geometric structure, but it is related with the properties of matter in a very direct way. So, it would be interesting to investigate how fluctuations in the matter content could destabilize the spacetime geometry, leading to, for example, a gyraton decaying in other configuration.

%========================================================
\subsection{What might work}
%========================================================
In some special situations, Griffiths has found quasi-classical type III$_0$ SETs based on massless Weyl fermions~\cite{Griffiths2}.
Specifically, the only concrete candidate to describe Bonnor's spinning null matter that we have found in the literature is that based on a quasi-classical limit of the massless Weyl neutrino field, as  proposed by Griffiths~\cite{Griffiths2}.
Indeed, one can note that the SET of the Weyl field presented in reference~\cite{Griffiths:1970ix} can be type III$_0$ under the necessary (not sufficient) condition of having a covariantly constant null vector $\nabla_a \ell_b=0$.

Therefore, some particular solutions of the EOM coming from Weyl Lagrangian could represent the matter content necessary to generate type III$_0$ spacetimes under the restriction of having $\nabla_a\ell_b=0$, (and not just $\nabla_a\ell_b\propto\ell_a\ell_b$).
It should be emphasized, however, that a \emph{physical} neutrino (with a non-vanishing mass) is now known to not be described by that massless Weyl Lagrangian. So, we do not expect the corresponding SET to be type III$_0$, due to the already mentioned instability of type III tensors under small perturbations. 

More importantly, it should be noted that fermions are  intrinsically non-classical particles that satisfy the Pauli exclusion principle. The potential problem here is not related with a possible quantum origin of type III$_0$ tensors, 
(note that type IV SET examples are already related to the consideration of quantum effects), but instead with extending a fermionic Lagrangian with Grasmannian-valued spinor fields to macroscopic gravitational scenarios.
Whereas it is clear how to consider a macroscopic description of bosonic particles by studying the corresponding classical field, it is not at all obvious how fermionic particles could be combined in a macroscopic field configuration satisfying the same properties (such as the same intrinsic angular momentum).
So, despite this interesting example provided by Griffiths~\cite{Griffiths2}, we prefer to continue our search of fundamental descriptions of macroscopic type III$_0$ SETs.

%========================================================
\subsection{What does work}
%========================================================
Start by noting that a type III$_0$ SET can be written as \cite{Martin-Moruno:essential, Martin-Moruno:III_0}
\begin{equation}
T_{ab}=f(\ell_a s_b+ s_a\ell_b),
\end{equation}
with $\ell^2=0$, $s^2=1$, and $\ell\cdot s=0$, where we have indentified the null vector appearing in the definiton of the type III SET, that is equation (\ref{typeIIISET}), with the geodetic, expansion-free, shear-free and twist-free vector of Kundt's geometries.
Although this is not necessarily always the case, this is useful to find particular examples.
%========================================================
\subsubsection{Ansatz based on using a non-dynamical background zero-divergence null vector}
%========================================================
%\enlargethispage{30pt}
Let us suppose that (as in all the examples 1 to 6 above) we have a zero divergence null vector field, $\ell^a$, which will we take to be a non-dynamical background field, and define the following Lagrangian density:
\begin{equation}
   L =     \ell^a  (\alpha \nabla_a \beta - \beta \nabla_a \alpha).
\end{equation}
Note that even in the most general case above we have $\nabla_a \ell_b\propto \ell_a \ell_b$, which implies $\nabla_a \ell^a = 0$.
The equations of motion for $\alpha$ and $\beta$ are then
\begin{equation}
\ell^a \nabla_a \beta + \nabla_a (\ell^a \beta) = 0  \qquad \Longrightarrow  \qquad \ell^a \nabla_a \beta = 0,
\end{equation}
and
\begin{equation}
  -\ell^a \nabla_a \alpha - \nabla_a (\ell^a \alpha) = 0 \qquad  \Longrightarrow \qquad  \ell^a \nabla_a \alpha = 0,
\end{equation}
where we have used the fact that $\ell^a$ has zero divergence.
Combining both equations of motion, we get
\begin{equation}\label{orto}
\ell^a  (\alpha \nabla_a \beta - \beta \nabla_a \alpha)    = 0,
\end{equation}
and, therefore, the Lagrangian density is zero on-shell: $L=0$. Consequently the (on shell) SET can be written as
\begin{equation}
 T_{ab} = \ell_a  (\alpha \nabla_b \beta - \beta \nabla_b \alpha) + \ell_b (\alpha \nabla_a \beta - \beta \nabla_a \alpha).
\end{equation}
Now define
\begin{equation}
s_a=\alpha \nabla_a \beta - \beta \nabla_a \alpha.
\end{equation}
Since $\ell^a$ is null, this SET is type III$_0$ as long as the vector $s_a$ is spacelike.
Now note that we already have $\ell^as_a=0$ in equation (\ref{orto}).
In all of the spacetimes we considered in the previous section $\ell^a = (\partial_r)^a$, so $\alpha$ and $\beta$ are functions of $(u,\,x,\,y)$. 
Finally, it can easily be checked that $g^{ab}s_a s_b=g^{ij}s_is_j$ for metric (\ref{general}). Since $g^{ij}$ is by construction nonsingular and Euclidean signature, $s_a$ is always spacelike as long as at least one of the components $s_i$ is nonzero.
While this Lagrangian is admittedly somewhat unusual, in particular $\ell$ is taken to be a non-dynamical externally imposed null vector field,  it certainly does generate a type~III SET.

%========================================================
\subsubsection{Background independent model}
%========================================================
Let us now consider the curved spacetime generalization of the Lagrangian that we first investigated for flat Minkowski space in reference \cite{Martin-Moruno:III_0}. This is
\begin{equation}
 L = F(\nabla_a A^a).
\end{equation}
The equation of motion of this Lagrangian is   
\begin{equation}
F''(\nabla \cdot A)\,\nabla_a (\nabla_b A^b) =0,
\end{equation}
where a prime denotes derivative with respect to the argument $\nabla \cdot A$.
Hence, for $F''\neq0$, the equation of motion implies that the scalar quantity $\nabla_b A^b$ is constant.
On the other hand, the SET of this Lagrangian is
\begin{equation}
    T_{ab} = \frac{1}{2} F'(\nabla A)  [\nabla_a A_b+ \nabla_b A_a] - \frac{1}{2} F(\nabla A) g_{ab}.
\end{equation}
Now, instead of specializing to Minkowski space, as we did in reference \cite{Martin-Moruno:essential}, we merely assume the existence of a null vector field such that
\begin{equation}
  \nabla_a \ell_ b = (\partial_r \Phi) \, \ell_a\ell_b.
\end{equation}
Such a vector field exists for all of the spacetimes we have considered herein whenever $\partial_r^2\Phi=0$ in metric (\ref{general}). Indeed in these situations $\ell^a = \partial_r$.
We now consider an \emph{ansatz} of the form 
\begin{equation}\label{ansatz}
A_a(x^\mu) = \ell_a(x^\mu)\; \chi(x^\mu),
\end{equation}
where $\chi(x^\mu)$ is at this point an arbitrary function.
Then 
\begin{equation}
\nabla_a A_b = (\partial_r \Phi) \, \ell_a\ell_b\,\chi + \ell_b \nabla_a \chi,
\end{equation}
and we easily see $\nabla_a A^a  = \ell^a \partial_a \chi$.
Therefore, a particular solution to the EOM  $\nabla_b A^b = {\rm (constant)}$    is      $\ell^a \, \partial_a \chi = 0$. 
But since   $\ell^a = (\partial_r)^a$   this simply implies   $\chi = \chi(u,\,x,\,y)$.
Therefore, the \emph{ansatz} (\ref{ansatz}) is indeed a solution on the equation of motion for  $\chi = \chi(u,\,x,\,y)$.
 As in the previous case, one can check that $g^{ab}\,\nabla_a \,\chi\nabla_b\chi = g^{ij} \,\partial_i \chi\,\partial_j\chi$, so   $\nabla_a\chi$ is spacelike for metric (\ref{general}) as long as at least one of the $\partial_i \chi$ is nonzero.
 The gradient $\nabla\chi$ is also orthogonal to $\ell$ since the solution of the EOM is $\ell^a\nabla_a \chi=0$. Then, for this solution to the EOM we have the following SET:
\begin{equation}
     T_{ab} = \frac{1}{2} F'(0)  \left\{ 2\, (\partial_r \Phi) \, \ell_a\ell_b\,\chi  +\ell_a \nabla_b \chi + \ell_b \nabla_a \chi\right\} -\frac{1}{2} F(0) g_{ab}.
\end{equation}
Now define
\begin{equation}
s_a = \nabla_a \chi +  (\partial_r \Phi) \, \chi\, \ell_a,
\end{equation}
and check that $s\cdot \ell=0$ and $s\cdot s=\nabla\chi\cdot\nabla\chi>0$. Then we can write
\begin{equation}
     T_{ab} = \frac{1}{2} F'(0)  [\ell_a s_b + s_a \ell_b] -\frac{1}{2} F(0) g_{ab},
\end{equation}
which is general type~III.
If we now choose the function $F$ such that $F(0)=0$, then 
\begin{equation}
      T_{ab} = \frac{1}{2}F'(0)  [\ell_a s_b + s_a \ell_b],
\end{equation}
which is type III$_0$.
While this Lagrangian is admittedly somewhat unusual, it does generate a type~III SET.

%========================================================
\section{Conclusions}\label{sec:conclusions}
%========================================================

Type~III SETs are the most unusual of the Hawking--Ellis classification. Finding a fundamental Lagrangian formulation leading to this type of tensor has long been somewhat of a mystery. Moreover, the physical interest of investigating this elusive type of SET is somewhat unclear if one does not know any physical situation that should be described by tensors of this type. In this work, we have reported significant progress in alleviating this situation.

We have focused our attention on the essential core of type~III tensors, the type~III$_0$ tensors. Reviewing the results regarding the only known example of type III$_0$ SET, the gyraton, we have been able to gain some intuition regarding type III$_0$ geometries. We have noted that the more basic gyraton examples are a sub-class of the Kundt geometries, which are geometries having a geodesic null vector with vanishing optical scalars, with the null vector being covariantly constant. The existence of a covariantly constant null vector was also a characteristic of the type III$_0$ spacetimes that we presented in reference~\cite{Martin-Moruno:III_0}.
However, we have soon noted that a more general set of type III$_0$ geometries can be found as a subclass of Kundt geometries requiring only that the null vector has zero divergence (due to the fact that $\nabla_a\ell_b\propto \ell_a\ell_b$).

As we are restricting our attention to type~III$_0$ tensors, we are considering only spacetimes with $R=0$. 
The reason why we have considered Kundt geometries and not the more general Robinson-Trautman spacetimes is that it is a subclass of Kundt geometries that having vanishing scalars invariants.
However, it may well be that Robinson-Trautman spacetimes have more general type~III SETs, or even type~III$_0$ tensors with any scalar invariant different from $R$ non-vanishing.

Regarding the fundamental Lagrangian formulation for the matter associated to type III SET, we have proven that it cannot just be the Maxwell Lagrangian. Some additional ingredient is necessary to avoid constraints coming from the Rainich condition which ensures that the electromagnetic SET is type I or II. 
We have also briefly discussed the construction carried out by Griffiths in terms of massless Weyl spinors, and indicated the limitations of that proposal.

Furthermore, we have presented two explicit classical Lagrangians leading to a type III$_0$ SET in curved spacetime. Although the construction of these Lagrangians can be considered in some sense artificial, in particular the first Lagrangian requires the existence of a  non-dynamical background null vector, they are to the best of our knowledge the first such examples presented in the literature. 

Therefore, although type III$_0$ SETs do not generically appear in classical or semi-classical situations, there are nevertheless examples of type III$_0$ spacetimes that are not so odd as might have been be expected.
The matter content necessary to support this type of SET does not seem to crop up in nature, at least not in any obvious form, but a consistent Lagrangian formulation can nevertheless be found. It is high time that we take care of the ugly duckling of the Hawking--Ellis classification, to conclude once and for all whether it is physically relevant or whether we can safely neglect it.

%========================================================
\section*{Acknowledgments}
%========================================================
PMM acknowledges financial support from the project FIS2016-78859-P (AEI/FEDER, UE). MV acknowledges financial support via the Marsden Fund administered by the Royal Society of New Zealand.

\appendix

%========================================================
\section{Comments on the Kundt class}\label{sec:ap}
\addcontentsline{toc}{section}{Appendix: Comments on the Kundt class}
%========================================================
Let us consider the spacetime geometry specified by
\begin{equation}\label{Kundt-more}
\hspace{-1.5cm} ds^2= -2 du \left[dr-H(u,\,r,\,x^k)\,du-W_i(u,\, r,\,x^k)\,dx^i\right]+g_{ij}(u,\,r, x^k)\,dx^idx^j.
\end{equation}
This is slightly \emph{more} general than what is defined as Kundt class in references~\cite{Stephani,Griffiths,Podolsky:2008,Coley:2009}, the  Kundt class corresponding to the restriction $g_{ij}(u,\,r, x^k)\to g_{ij}(u, x^k)$.
Indeed, metric (\ref{Kundt-more}) describes a $4$-dimensional Robinson-Trautman spacetime \cite{Podolsky:2018oov}, which reduces to Kundt geometry when the expansion of the congruence associated with the null geodetic vector vanishes.
Temporarily retaining the full generality of (\ref{Kundt-more}), the null vector $\ell=\partial_r$ can be written as
\begin{equation}
\ell^a=(0,1,0,0)\qquad\hbox{so} \qquad \ell_a=(-1,0,0,0).
\end{equation}
A brief computation yields $\ell^b \nabla_b \ell^a=0$, so $\ell$ is in fact an affinely parameterized null geodesic vector field. 
Furthermore, one can easily verify that
\begin{equation}\label{E:covl1}
\nabla_a \ell_b =  {1\over2} \partial_r g_{ab}.
\end{equation}
Therefore, $\ell$ will be a Killing vector (meaning $\nabla_{(a}\ell_{b)}=0$) if and only if it is covariantly constant; that is, $\nabla_a \ell_b=0$. This will happen if and only if the functions in metric (\ref{Kundt-more}) do not depend on the spacelike coordinate $r$. Note that in Kundt geometries we have  a geodesic null congruence with vanishing optical scalars, but the tangent vector does not necessarily need to be Killing.

One can explicitly check this by considering the metric
\begin{equation}\label{Kundt-less}
 ds^2= -2 du \left[dr-H(u,\,x^k)\,du-W_i(u,\,x^k)\,dx^i\right]+g_{ij}(u,\, x^k)\,dx^idx^j,
\end{equation}
which is slightly \emph{less} general than what is called Kundt class in references~\cite{Stephani,Griffiths,Podolsky:2008,Coley:2009}.
Another brief calculation yields
\begin{equation}\label{E:covl2}
\nabla_a \ell_b =  0.
\end{equation}
So in this situation, (which is the one we are actually most interested in for type~III$_0$ SETs), $\ell^a$ is simultaneously a null covariantly constant vector field \emph{and} a Killing vector.

Finally, we can also use expression (\ref{E:covl1}) to easily check that we only need $\partial_rg_{ij}=0$ to have a Kundt spacetime. Starting from the more-general-than-Kundt metric (\ref{Kundt-more}), the congruence of null geodesics with tangent vector $\ell$ has expansion tensor, vorticity tensor, and shear tensor given by
\begin{equation}
\theta_{ij}=\nabla_i \ell_j =  \frac{1}{2}\partial_r g_{ij},\quad \omega_{ij}=\nabla_{[i} \ell_{j]}=0,\quad\sigma_{ij}=\theta_{ij}-\frac{1}{2}\theta \,h_{ij},
\end{equation}
respectively, where $\theta=\theta^i{}_i$ is the expansion scalar, $h_{ij}$ is the metric induced in the spatial 2-surfaces, and we have taken into account equation (\ref{E:covl1}). So, these tensors and the associated scalars vanish if $\partial_r g_{ij}=0$.  
That is, the Kundt class corresponds to
\begin{equation}\label{Kundt-exact}
\hspace{-1cm} ds^2= - 2 du \left[dr-H(u,\,r,\,x^k)\,du-W_i(u,\, r,\,x^k)\,dx^i\right]+g_{ij}(u,\, x^k)\,dx^idx^j.
\end{equation}
Once we are working within the Kundt class of (\ref{Kundt-exact}) or equivalently (\ref{Kundt}) we have
\begin{equation}
\nabla_a \ell_b =\frac{1}{2} \left[ \begin{array}{c|c|c} 2\partial_r H & 0 & \partial_r W_j \\ \hline 0 & 0 & 0\\ \hline \partial_r W_i & 0 & 0\end{array} \right].
\end{equation}
While this is not a SET, merely a symmetric $T^0_2$ tensor, nothing prevents us from applying the purely \emph{algebraic} 
aspects of the Hawking--Ellis classification to this object  so that
$\nabla_a \ell_b = - \nabla_a \nabla_b u$ (note that $\ell_a = (-1,0,0,0)  = -\partial_a u)$, and to consequently deduce that $\nabla_a \ell_b = - \nabla_a \nabla_b u$ is \emph{algebraically} a type III$_0$ tensor for the entire Kundt class of spacetimes.

In summary, in order to have a Kundt geometry we do not need  a covariantly constant null Killing vector, it is sufficient to have a null geodesic vector such that $\nabla_i\ell_j=0$. 
It should be emphasized that all the type III$_0$ spacetimes we have studied above, in addition to having $\nabla_i\ell_j=0$, also satisfy $\nabla_a\ell_b\propto(\partial_rH)\,\ell_a\ell_b$, where $\partial_rH$ might vanish or not. 
Therefore, those spacetimes belong to a subclass of Kundt spacetimes for which the geodesic null vector is divergence-free, that is $\nabla_a\ell^a=0$, although it is not necessarily covariantly constant.

\section*{References}
%%%%%%%%%%%%%%%%%%%%%%%%%%%%%%%%%%%%%
%%%%%%%%%%%%%%%%%%%%%%%%%%%%%%%%%%%%%

%
\end{document}